\documentclass[10pt]{iopart}
\usepackage{geometry}                
\usepackage[parfill]{parskip}    
\usepackage{graphicx}
\usepackage{amssymb}
\usepackage{epsfig}
\usepackage{epsf}
\usepackage{epstopdf}
\usepackage[hypertex]{hyperref}

\begin{document}
\title{Dipole diffusion in a random electrical potential}
\author{Cl\'ement Touya, David S. Dean, and Cl\'ement Sire}

\address{Laboratoire de Physique Th\'eorique -- IRSAMC,
Universit\'e de Toulouse, CNRS, 31062 Toulouse, France}

\begin{abstract}
We study the Langevin dynamics of a  dipole diffusing in a random
electrical field ${\bf E}$ derived from a quenched Gaussian
potential. We show that in a suitable adiabatic limit (where the
dynamics of the dipole moment is much faster than the dynamics of
its position), one can reduce the coupled stochastic equations to an
effective Langevin equation for a particle diffusing in an effective
potential with a spatially varying and anisotropic local diffusivity
$\kappa_{ij}$. Analytic results, close to the adiabatic limit, for
the  diffusion constant $\kappa_e$ are found in one dimension and a
finite temperature dynamical transition is  found. The system is
also studied numerically. In particular, we study the anomalous
diffusion exponent in the low temperature regime. Our findings
strongly support the conclusion that the location of the dynamical
transition and the anomalous diffusion exponents are  determined by
purely static considerations, {\em i.e.} they are independent of the
relative values of the diffusion constants of the particle position
and its dipole moment.

\end{abstract}
\maketitle

\section{Introduction}

Computing bulk transport properties in random media is an important
physical problem \cite{rev,ddh,dech,mathe,king1,king2,dewit,spos}
having many applications. Two important models of random media have
been widely studied (i) where the randomness is due to a random
potential and (ii) where the local diffusions constant is random.
There are mathematical relations between the problems (i) and (ii)
and problems of type (ii) are also related to a wide variety of
physical problems such as the effective conductivity and dielectric
constants of random conductors and dielectrics and the effective
permeability of porous media \cite{rev}. An important quantity to
understand those properties is the {\it late time diffusion
constant} of a Brownian tracer particle in interaction with the
medium. In the system we will study, the local transport properties
are modified by the interaction of the dipole moment of the tracer
particle with a random electric field  drawn from an appropriate
statistical ensemble. If the electric field disorder is
statistically invariant under translation and short range correlated
in space, we expect that the diffusion constant and mobility due to
an applied force are self-averaging in the regime where the
diffusion is normal. If the  transport is normal, then in the long
time limit ($t\to\infty$), the mean squared displacement of the
particle position  behaves as

\begin{equation}
\langle({\bf x}(t)-{\bf x}(0))^2\rangle\sim 2D\kappa_et\label{meandisp},
\end{equation}

where $D$ is the dimension of space and $\kappa_e$ is the late time
diffusion or effective diffusion constant. In general there are
subdominant corrections to the above, which depend strongly on the
dimensionality of the problem \cite{bouch,tode}.

If  the random field is  frozen or evolves over  very long time
scales with respect to those of the tracer particle, then one is in
the situation of  {\it quenched disorder}, where the potential does
not depend on time. The motivation for studying models of particles
diffusing in a random potential comes from the fact that such
systems arise very naturally in nature; for instance in zeolites
where random electric fields are generated by the presence of frozen
charged impurities \cite{chen}. If the particle has a net charge,
then the particle interacts with the electric field and the case of
diffusion in a quenched scalar field has been extensively studied in
the literature \cite{rev,ddh,dech,tode,deto}. However, if the
particle is polarizable but without charge, it also interacts with
the field. In this case, the problem is quite different and much
less well studied \cite{deto,dhs}. We also note that systems  with
quenched disorder, spin glasses for example,  are often good
paradigms for systems having structural glass transition (where no
random field is present). It is often argued heuristically, that for
sufficiently complex and frustrated systems, a single particle in
the system sees an effectively  random potential due to the other
particles. At a mean field level, there exist models, where this
analogy has been used successfully to analyze the statistical
mechanics of frustrated but non-disordered models. In practice one
can have two models, one with quenched disorder and the other
without but highly frustrated, which exhibit the same thermodynamics
in the high temperature phase and the same glass transition at low
temperatures \cite{mari1,mari2,bouchmez}. Even if the frustrated
non-random system possesses a crystalline ground state, not shared
by the disordered system, this fact is practically irrelevant as
this state is dynamically never attained. Therefore, these models
are often adopted as toy models for structural glass transition. In
the liquid phase, we expect the tracer particle to have a non-zero
diffusion constant $\kappa_l$. Now consider the situation in which
the same particle diffuses in a quenched background, where all the
other particles have been frozen in a particular configuration. A
realistic choice would be to select a configuration from a
Gibbs-Boltzmann equilibrium ensemble. If $\kappa_e$ is the diffusion
constant, it has been shown \cite{demasi,osada} that
$\kappa_e<\kappa_l$ which makes physical sense, as if the background
particles can move about, the cages, which  trap the tracer , will
break up on some time scale and free it to disperse more quickly
than in the quenched case. Moreover, in some special cases,
$\kappa_e$ can vanish at a critical temperature or disorder strength
\cite{tode,deto}. If the system has a finite correlation length
$l_0$ (which will be the case, as our field will be short range
correlated), the diffusion constant can be used to give an effective
relaxation time $\tau$

\begin{equation}
\tau\sim\frac{l_0^2}{\kappa_e}.\label{tau}
\end{equation}

If $\kappa_e\to 0$, it means, that we have a diverging timescale in the same way in
which a structural glass has a diverging timescale which, experimentally, can be
extracted from the divergence of the liquid's viscosity.

In this paper we study the physical case of dipoles diffusing
in a random electric field ${\bf E}({\bf x})$ which is spatially varying
though time independent (or quenched). It is generated by a random
potential $\phi({\bf x})$ which gives
${\bf E}({\bf x})=-\nabla\phi({\bf x})$, with correlation function

\begin{equation}
\langle\phi({\bf x})\phi({\bf x}^\prime)\rangle=
\Delta(|{\bf x}-{\bf x}^\prime|).\label{phicorr}
\end{equation}

The potential $\phi({\bf x})$ is thus statistically isotropic and
invariant by translation in space. The most convenient
choice is to take $\phi$ to be Gaussian. If the dipole moment,
denoted by ${\bf p}$, is modelled as two opposite charges
connected to a Harmonic spring and ${\bf x}$ denotes the position of the
dipole centre, then the total energy of a particle at the point
(${\bf p}$, ${\bf x}$) is given by

\begin{equation}
H({\bf x},{\bf p})={1\over 2\chi}{\bf p}^2 -{\bf p}\cdot {\bf E}({\bf x})\label{eqH}.
\end{equation}

The first term is the Harmonic energy of the spring, and $\chi$ is
simply the dipole polarisability. The second term is the energy of
the dipole's interaction with the field ${\bf E}$. The partition
function for the system, where ${\bf x}$ is confined to a volume
denoted by $\cal{V}$ of a $D$ dimensional space, is given, up to an
overall factor, by

\begin{equation}
Z =\int_{\cal{V}}d{\bf x}\int_{\mathbb R^D}d{\bf p}\exp\left(-\beta{{\bf p}^2\over 2\chi }
+\beta{\bf p}\cdot {\bf E}({\bf x})\right),\label{part}
\end{equation}
where $\beta=1/k_BT$ is the inverse temperature.
If we trace over the dependence on ${\bf p}$, we find an effective
partition function for the variable ${\bf x}$ given, again up to a constant, by

\begin{equation}
Z_{eff} = \int_{\cal {V}} d{\bf x} \exp\left(-\beta V({\bf x})\right),\label{parteff}
\end{equation}

where $V$ is the effective potential for ${\bf x}$ and is given by

\begin{equation}
V({\bf x}) = - {\chi {\bf E}^2({\bf x})\over 2 } \label{eqV}.
\end{equation}

This represents the case where the dipole moment adapt instantaneously to
the external field and motivated the study in \cite{tode,deto}
 of diffusion in non-Gaussian potentials. In \cite{tode} the diffusion
 constant for a particle diffusing in non Gaussian potentials such as
 given by equation (\ref{eqV}) was computed exactly in one dimension,
and in \cite{deto}  a self renormalization group scheme was developed to study the problem
in higher dimensions.

In \cite{tode,deto} a critical temperature at which the diffusion
constant vanishes, signalling  a dynamical transition from a normal
diffusive regime to a subdiffusive regime was identified. We thus
expect that, in a suitable adiabatic limit where the dynamics of the
dipole degree of freedom ${\bf p}$ is much more rapid than that of
the spatial variable ${\bf x}$, that the effective potential seen by
the variable ${\bf x}$ is $V$ as defined by equation (\ref{eqV}).

If we take an overdamped
Langevin dynamics for both the dipole and positional degrees of freedom, the
equations of motion are given by

\begin{eqnarray}
{dx_i\over dt} &=& \beta \kappa_x p_j {\partial E_j \over \partial x_i}
+\sqrt{2 \kappa_x}\eta_{x_i}  \label{leqx} \\
{dp_i\over dt} &=& -\kappa_p \beta\left({p_i\over \chi} - E_i\right)
+ \sqrt{2\kappa_p} \eta_{p_i}. \label{leqp}
\end{eqnarray}

In the above equations $\kappa_x$ is the bare diffusion constant for the
spatial variable  and the other diffusion constant
$\kappa_p$ sets the time scale for the relaxation of the dipole and the
adiabatic limit is  where $\kappa_p\to\infty$. In the dynamical  equations
(\ref{leqx}) and (\ref{leqp}), terms of the form  $A_i$ represent the component
of the vector ${\bf A}$ in the direction $i$ ; we have also used the Einstein
summation convention and will stick with this convention throughout the paper.
The noise terms are white noise and their correlation functions are given by

\begin{equation}
\langle \eta_{p_i} (t)\eta_{p_j} (s)\rangle = \langle \eta_{x_i}
 (t)\eta_{x_j} (s)\rangle = \delta_{ij}\delta(t-s)\textrm{ and
 }\langle \eta_{p_i} (t)\eta_{x_j} (s)\rangle = 0.\label{eqwn}
\end{equation}

If we rewrite (\ref{leqp}) and substitute it  into equation  (\ref{leqx}) we find

\begin{equation}
{d{x}_i\over dt} = \sqrt{2\kappa_x}{\bf
\eta}_{x_i}+\kappa_x\beta{\partial \over \partial x_i} \frac{\chi
E_j^2}{2}-\frac{\kappa_x}{\kappa_p}\chi{\partial E_j \over \partial
x_i} \left({dp_j\over dt}-\sqrt{2\kappa_p}{\bf
\eta}_{p_j}\right)\label{adia},
\end{equation}

If we now take the adiabatic limit ($\kappa_p \gg\kappa_x$),
equation (\ref{adia}) reduces to the Langevin equation for a
particle in the potential $V({\bf x})$ defined by equation
(\ref{eqV}). In appendix A, we will rederive this result in a more
rigorous way and also see, at first order, the effect of a finite
value of $\kappa_p$.  We thus expect, that for very large values of
$\kappa_p$, we find the same dynamics for ${\bf x}$ studied in
\cite{tode,deto,dhs}.  Consequently there should also be  a
dynamical transition in this problem when the field $\phi$ is
Gaussian. We can now ask the question, what happens if the time
scale for the relaxation of the dipole moment is non zero? Will the
dynamical transition remain or is it
a pathology of the limit $\kappa_p\to\infty$?\\
In the appendix A we derive the effective dynamics for the marginal
distribution of ${\bf x}$   to order  $\kappa_x/\kappa_p$. To this
order the effective dynamics of ${\bf x}$ still can be described by
a Langevin equation, with the same potential as equation (\ref{eqV})
but with a spatially varying non-isotropic diffusion constant. This
is a rather remarkable fact and can be shown using operator
projection  techniques \cite{risk}. However we will present a
derivation based on a direct manipulation  of the Langevin equations
similar to that of  \cite{sasa}. We chose this route as it gives a
physical feeling for why the effective process for ${\bf x}$ is to
this order Markovian and also because the computation in spaces of
dimension greater than one are more straightforward within this
formalism. We also compute the first order corrections to the
effective diffusion constant in the high temperature limit via a
Kubo formula for the effective diffusion constant.

In section (3) we will present exact result for the diffusion
constant in one dimension. In section (4) we will confront our
results with numerical simulation (stochastic second order
Runge-Kutta) of the coupled Langevin equations (\ref{leqx}) and
(\ref{leqp}) in the diffusive and sub-diffusive regime. Finally in
section 5 we will conclude and discuss our results.

\section{Large $\kappa_p$ and small $\beta$ approximations}

The time scale for the relaxation of the dipoles should be
proportional to $\kappa_p^{-1}$ and thus, the adiabatic limit (where
the dipoles adapt very quickly to the local field) will correspond
to the limit where $\kappa_p$ becomes large. Interestingly, in this
limit, one can simplify the coupled equations (\ref{leqx}) and
(\ref{leqp}) to an effective Langevin equation for ${\bf x}$ up to
the order $O(\kappa_x/\kappa_p)$ by direct manipulation of the
Langevin equations \cite{sasa}. The exact derivation is a little
long and technical and for clarity's sake it is thus given in
appendix A. As a result, we can write down the following
Fokker-Planck equation which describe the effective process for
${\bf x}$

\begin{equation}
{\partial \rho\over \partial t}= H \rho = {\partial \over \partial
x_k}\left[\kappa_{ki}\left({\partial \rho\over \partial x_i} + \beta \rho
{\partial V\over \partial x_i} \right)\right].\label{hfinal}
\end{equation}

Where $V$ is the effective potential given by equation ({\ref{eqV})
and $\kappa_{ij}$ is a spatially varying and anisotropic diffusivity
tensor given by

\begin{equation}
\kappa_{ij} = \kappa_x\left[ \delta_{ij} - {\kappa_x\over \kappa_p}
\chi^2 {\partial E_k\over \partial x_i}{\partial E_k\over
\partial x_j}\right].\label{kij}
\end{equation}

The Fokker-Planck equation (\ref{hfinal}) has the correct Gibbs
Boltzmann equilibrium distribution with the effective
potential $V $ and from this one can write down a Langevin equation
which corresponds to the process

\begin{equation}
{dx_i\over dt} = -\beta{\partial \over \partial x_i}\frac{\chi E_j^2}{2}
+{\partial \over \partial x_i}\kappa_{ij}+\sqrt{2\kappa_{ij}}\eta_i.
\label{eflang}
\end{equation}

We note that if we take the limit where $\kappa_p\to\infty$ in
(\ref{eflang}), the diffusivity reduce to
$\kappa_{ij}=\kappa_x\delta_{ij}$ and we recover the Langevin
equation for a particle in the potential $V({\bf x})$ rigorously.
In equation (\ref{opeexp}) of appendix A we performed an expansion
in $\alpha^{-1}$ (with $\alpha=\kappa_p\beta/\chi$) and assumed
$\alpha$ to be large. However if $\kappa_p$ is large but finite we
see that  as $\beta$ becomes small (the high temperature limit) the
expansion will fail. To predict the behavior of $\kappa_e$ at high
temperature  we will thus use a Kubo formula for the effective
diffusion constant. Integrating the stochastic differential equation
(\ref{leqx}) between $0$ and $t$ we have

\begin{equation}
x_i(t) - x_i(0)=\sqrt{2\kappa_x}B_i(t)+\beta
\kappa_x\int_0^tdsp_j(s).\nabla_iE_j({\bf x}(s))
\end{equation}

Where ${\bf B}_t$ is a standard $D$-dimensional Brownian motion with
$\langle{\bf B}_t^2\rangle=2Dt$ and $x_i(0)$ is the initial position
in the direction $i$. Thus squaring the above equation and taking
the average yields

\begin{eqnarray}
\displaystyle
2\kappa_xt=\langle\left(x_i(t)-x_i(0)\right)^2\rangle+
2\kappa_x\beta\langle\left(x_i(t)-x_i(0)\right).
\int_0^tdsp_j(s).\nabla_iE_j({\bf x}(s))\rangle\nonumber\\
\displaystyle
+(\kappa_x\beta)^2\int_0^tds\int_0^sds^\prime\langle p_j(s).
\nabla_iE_j({\bf x}(s))p_k(s^\prime).\nabla_iE_k({\bf x}
(s^\prime)\rangle.\label{xintsq}
\end{eqnarray}

Using the property of detail balance, which ensures time
translation invariance at equilibrium, we have for any
two functions $A$ and $B$ that

\begin{equation}
\langle A({\bf x}(t))B({\bf x}(s))\rangle=\langle
A({\bf x}(t-s))B({\bf x}(0))\rangle,\label{symet}
\end{equation}

and the Onsager symmetry relation

\begin{equation}
\langle A({\bf x}(t))B({\bf x}(s))\rangle=\langle
B({\bf x}(t))A({\bf x}(s))\rangle.\label{onsager}
\end{equation}

Assuming that the system starts in equilibrium \cite{comment}  one
can thus apply (\ref{symet}) and (\ref{onsager}) to the second term
of (\ref{xintsq}) and show that it should vanish

\begin{eqnarray}
\displaystyle
\langle\left(x_i(t)-x_i(0)\right).\int_0^tdsp_j(s).
\nabla_iE_j({\bf x}(s))\rangle\nonumber\\
\displaystyle
=\int_0^tds\left[\langle x_i(t)p_j(s).\nabla_iE_j({\bf x}(s))
\rangle-\langle x_i(0)p_j(s).\nabla_iE_j({\bf x}(s))
\rangle\right]\nonumber\\
\displaystyle
=\int_0^tds\left[\langle x_i(t-s)p_j(0).
\nabla_iE_j({\bf x}(0))\rangle-\langle x_i(0)p_j(s).
\nabla_iE_j({\bf x}(s))\rangle\right]\nonumber\\
\displaystyle
=\int_0^tds\left[\langle p_j(s).\nabla_iE_j({\bf x}(s))
x_i(0)\rangle-\langle x_i(0)p_j(s).\nabla_iE_j
({\bf x}(s))\rangle\right]\nonumber\\
=0.
\end{eqnarray}

Thus, (\ref{xintsq}) reduces to

\begin{eqnarray}
\langle\left(x_i(t)-x_i(0)\right)^2\rangle=2\kappa_xt\nonumber\\
-(\kappa_x\beta)^2\int_0^tds\int_0^sds^\prime
\left<p_j(s-s^\prime)p_k(0)\nabla_iE_j
({\bf x}(s-s^\prime))\nabla_iE_k({\bf x}(0))\right>.\label{kex1}
\end{eqnarray}

This equation is  however exact  and can be evaluated to order ${\bf
E}^2$ by calculating the integral  on its right hand side using the
statistics for ${\bf x}$ and ${\bf p}$ in the absence of ${\bf E}$,
{\em i.e.} in the weak disorder limit which should become exact at
high temperatures. Here ${\bf p}$ and ${\bf x}$  are purely Gaussian
with correlation functions
\begin{equation}
\left<p_j(s)p_k(0)\right>=\frac{\chi}{\beta}
\exp\left(-\frac{\kappa_p\beta}{\chi}s\right)\delta_{jk}
,\label{pertcorp}
\end{equation}
and
\begin{equation}
\langle\left({x}_i(t)-{x}_j(0)\right)^2
\rangle =2\delta_{ij}\kappa_x t.
\end{equation}

This approximation yields

\begin{eqnarray}
\left<p_j(s-s^\prime)p_k(0)\nabla_iE_j
({\bf x}(s-s^\prime))\nabla_iE_k({\bf x}(0))\right>
\nonumber\\
=\frac{\chi}{\beta}\exp\left(-\frac{\kappa_p\beta}{\chi}(s-s^\prime)\right)
\langle\nabla_iE_j({\bf x}(s-s^\prime))
\nabla_iE_k({\bf x}(0))\rangle\delta_{jk}.
\end{eqnarray}

Summing over the spatial indices then gives

\begin{equation}
\kappa_e=\kappa_x-\frac{\kappa_x^2\beta\chi}{Dt}
\int_0^tds\int_0^sds^\prime\exp\left(-\frac{\kappa_p\beta}
{\chi}(s-s^\prime)\right)\langle\nabla_i{ E}_j({\bf x}(s-s^\prime))
\nabla_i{E}_j({\bf x}(0))\rangle.
\end{equation}

The correlation function of the random field component in the above
can be computed in terms of the correlation function $\Delta({\bf
x})$ of the random electrostatic field $\phi$.

\begin{eqnarray}
\left<(\nabla_i{ E}_j{\bf x}(s-s^\prime))\nabla_i{ E}_j({\bf x}(0))\right>&=&
\int\frac{d{\bf k}}{(2\pi)^D}{\tilde\Delta}({\bf k})
{\bf k}^4\left<\exp(i{\bf k}.{\bf x}(s-s^\prime))\right>\nonumber\\
&=&\int\frac{d{\bf k}}{(2\pi)^D}{\tilde\Delta}({\bf k})
{\bf k}^4\exp(-{\bf k}^2\kappa_x(s-s^\prime))\label{pertcorep}
\end{eqnarray}

where ${\tilde \Delta}({\bf k})$ is the Fourier
transform of $\Delta({\bf x})$. Finally, we are left with
a simple double integration over $s$ and $s^\prime$,
 and from the large time behaviour we can extract $\kappa_e$
 as

\begin{equation}
\kappa_e=\kappa_x-\frac{\kappa_x^2\beta\chi}{D}
\int\frac{d{\bf k}}{(2\pi)^D}\frac{{\tilde\Delta}
({\bf k}){\bf k}^4}{{\bf k}^2\kappa_x+
\frac{\kappa_p\beta}{\chi}}.\label{kubo}
\end{equation}

\section{Analytic results in one dimension}

In the previous section, all the results we derived were for an
arbitrary dimension $D$. To analyze our results to first order
beyond the adiabatic limit, we will restrict our study to the one
dimensional case where one can compute exactly the diffusion
constant $\kappa_e$ for the effective Fokker-Planck equation
(\ref{hfinal}). We apply the general results of \cite{rev} to
compute the effective diffusivity via a static problem:

\begin{equation}
\kappa(x)\left(\frac{d\rho}{dx}+\beta\rho\frac{dV}{dx}
\right)=j,\label{current}
\end{equation}

where $j$ is the current and $\kappa(x)$ is the spatially
varying diffusivity in one dimension

\begin{equation}
\kappa(x)=\kappa_x\left[1-{\kappa_x\over \kappa_p}\chi^2
\left(\frac{dE(x)}{dx}\right)^2\right].\label{diff1d}
\end{equation}

We can solve (\ref{current}) and find

\begin{equation}
\rho(x)=je^{-\beta V(x)}\int_0^x\frac{e^{\beta
V(y)}}{\kappa(y)}dy\label{rhox}
\end{equation}

The diffusion constant $\kappa_e$ is then given as
$\kappa_e\left<\frac{d\rho}{dx}\right>=j$. Where $\left<.\right>$
is the average over the disorder in the
random electrical field. This gives
\begin{equation}
\kappa_e\left<\frac{d\rho}{dx}\right>=\lim_{L\to\infty}\frac{\kappa_ej}{L}
\left<e^{-\beta V(L)}\int_0^L\frac{e^{\beta V(y)}}
{\kappa(y)}dy\right>=j,
\end{equation}
and finally
\begin{equation}
\kappa_e=\frac{1}{\left<e^{-\beta V}\right>
\left<\frac{e^{\beta V}}{\kappa}\right>}.\label{ke}
\end{equation}

This result can also be obtained
via a first passage time argument \cite{zwa,deg}.
The electrical potential $\phi(x)$ is Gaussian and
if  we choose a correlator of the form
$\Delta(x)=f(x^2)$, where $f$ is analytic at $x=0$, then we find that
$\left<E^2(0)\right>=-2f'(0)$,
and $\left<E^\prime(0)E(0)\right>=0$.
Therefore, $E^\prime$ and $E$ are  uncorrelated and we can write
\begin{equation}
\kappa_e=\frac{\left<\kappa^{-1}\right>^{-1}}{\left<e^{-\beta V}\right>
\left<{e^{\beta V}}\right>}.\label{ke1}
\end{equation}
In \cite{tode} (corresponding to the adiabatic case here) it was
shown that the dynamical transition could be identified, via an
Arrhenius type argument, with the divergence of one or the other of
the first  two terms on the denominator of the equation above. We
thus see that to first order beyond the adiabatic approximation the
location of the transition temperature should be the same in the
adiabatic limit and close to this limit.

In the numerical simulations we will carry out we take the choice of
correlation function $f(u)=\exp(-\frac{1}{2}u)$ for which we find

\begin{equation}
\kappa_e=\kappa_x(1-\chi^2\beta^2)^{1/2}(1-
\frac{\kappa_x\chi^2}{\kappa_p}).\label{ke1d}
\end{equation}

\section{Numerical simulations}

In this section we test our analytical predictions against numerical
simulations of the Langevin equations (\ref{leqx}) and (\ref{leqp})
in one dimension. In our simulation, we set the diffusion constant
without disorder $\kappa_x$ to $1$ for convenience and we used
rescaled variables $p_i={\tilde p}_i\sqrt{\kappa_p}$ and
$E_i={\tilde E}_i \sqrt{\kappa_p}$, so that both equations have the
same time scale,

\begin{eqnarray}
{dx_i\over dt} &=& \beta \kappa_p {\tilde p}_j {\partial
{\tilde E}_j \over \partial x_i}+\sqrt{2 }\eta_{x_i}\label{leqx2} \\
{d{\tilde p}_i\over dt} &=& -\kappa_p \beta\left({{\tilde p}_i
\over \chi} - {\tilde E}_i\right) + \sqrt{2} \eta_{p_i} \label{leqp2}
\end{eqnarray}

The coupled stochastic differential equations were integrated using
second order Runge-Kutta integration schemes developed in
\cite{dhh,hon} and reviewed in \cite{rev} with a time step $\Delta
t=0.001$. In all simulations the effective diffusion constant for a
given realization of the disorder was obtained by fitting the mean
squared displacement averaged over $2000$ particles at late times.
The time of the simulation was chosen so that particles had
typically diffused ten or so correlation lengths of the field. The
fit of the average mean square displacement was done over the last
half of the time of the simulation (to ensure that the mean squared
displacement is well within the linear regime) by a non linear form:
$at+bt^\theta$, with $\theta<1$. The correction to the linear term
has to be taken into account as it becomes more and more relevant
close to the transition due to the slowing down of the dynamics
\cite{tode}.

To generate the Gaussian field characterized by a correlator
(\ref{phicorr}), we used the technique presented in
\cite{masi,sire}. The process can be written in a general form

\begin{equation}
\phi(x)=\int_{-\infty}^{+\infty}K(x-x^\prime)\eta(x^\prime)\,dx^\prime
,\label{processphi}
\end{equation}

where $K$ is a kernel function and $\eta(x)$ is a Gaussian white
noise. The linear form of (\ref{processphi}), ensures that $\phi(x)$
is a Gaussian process and the translationary invariance of the
kernel $K(x-x^\prime)$ ensures its stationarity. Now taking the
Fourier transform defined by ${\tilde
\phi}(\omega)=\int_{-\infty}^{+\infty} \phi(x)\exp(-i\omega x),dx$
of (\ref{processphi}), we find

\begin{equation}
{\tilde\phi}(\omega)={\tilde K}(\omega){\tilde\eta}(\omega).\label{fourrx}
\end{equation}

If  we now take the Fourier transform of the correlator (\ref{phicorr})
and the correlator $\langle\phi(\omega)\phi(\omega^\prime)\rangle$ from
(\ref{fourrx}), then by identification, we have
${\tilde\Delta}(\omega)=|{\tilde K}(\omega)|^2$  and thus from (\ref{fourrx})

\begin{equation}
{\tilde\phi}(\omega)=\sqrt{{\tilde\Delta}(\omega)}{\tilde\eta}(\omega)
\end{equation}

We can now create $\phi(x)$ by sampling ${\tilde\phi}(\omega)$ on a
frequency mesh and taking the inverse fast Fourier transform (FFT).
In \cite{tode,deto}, the authors used a method due to Kraichnan
\cite{rev,kra} which gives some finite size corrections for the
diffusion constant close to the transition.  However as we wish to
precisely locate the transition and tests its dependence on
$\kappa_p$ (and as we do not have analytical results for all
parameter ranges) we use this FFT based method. Again we take a
correlator of the form $\Delta(x)=\exp(-\frac{1}{2}x^2)$ and we
average our sample over $2000$ realizations of the field. In all our
simulation we set $\chi=1$. We measured the average value of
$\kappa_e$ over the field with an error bar estimated from the
standard deviation from sample to sample. The result in one
dimension is compared in figure (\ref{kes}) for $\kappa_p=10$ with
the analytical result (\ref{ke1d}) and the Kubo formula
(\ref{kubo}). As expected, we are well in the adiabatic regime and
(\ref{ke1d}) is in a very good agreement with the numerical
simulations. Moreover, we find a dynamical transition at $\beta_c=1$
where the diffusion constant becomes zero and below the critical
temperature $T_c$, the diffusion turns out to be anomalous (see
later).   Finally, the agreement between (\ref{ke1d}) and the
numerical results breaks down at high temperature, but fortunately
in this regime the results agree with the weak disorder/high
temperature analytical expression (\ref{kubo}). We have thus a
crossover between two analytical approximation which allows us to
predict the behavior of $\kappa_e$ in the whole normal diffusion
regime.

\begin{figure}
\begin{center}
\epsfxsize=0.6\hsize\epsfbox{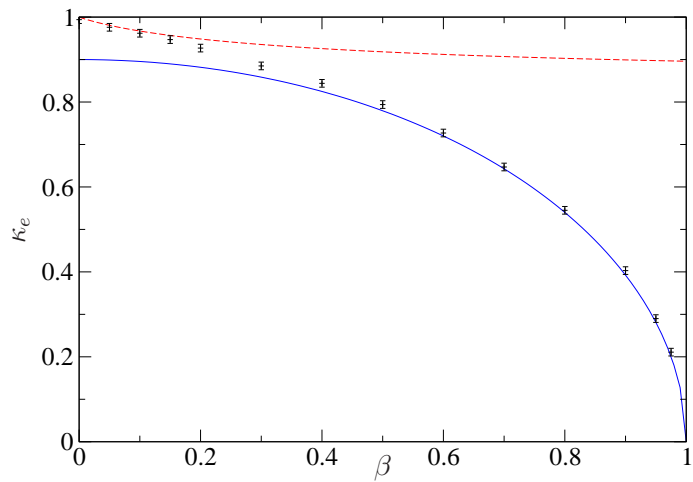}
\end{center}
\caption{Numerical value of $\kappa_e$ (stars) for $\kappa_p=10$,
compared with analytical results (\ref{kubo}) (dashed line) and
(\ref{ke1d}) (solid line).}
\label{kes}
\end{figure}

Beyond the adiabatic approximation, that is to say when $\kappa_p$
is of the same order of $\kappa_x$ or  smaller, we lack analytical
results (except for weak disorder), thus we must  calculate $\kappa_e$
numerically. The results are plotted  in figure (\ref{kp1}) for $\kappa_p=1$
 and figure (\ref{kp0.1}) for $\kappa_p=0.1$ and compared with the weak disorder
 result of equation (\ref{kubo}). In both cases, the transition
still appears to occur   at, or very close to,  the same critical
temperature $T_c=1$. Moreover, the diffusion constant clearly
decreases monotonically with the value of $\kappa_p$ at fixed
temperature. The low temperature phase, below $T_c$, is
characterized by an anomalous sub-diffusive behavior

\begin{equation}
\langle({\bf x}(t)-{\bf x}(0))^2\rangle\sim Ct^{2\nu},\label{meandispa}
\end{equation}

where the exponent associated with the anomalous diffusion
$\nu<1/2$. The subject of anomalous diffusion in disordered
media has been extensively studied  and a good review of it can be found in  \cite{bouch}.
In the case of $\kappa_p\to\infty$ the authors in \cite{tode}
were able to evaluate $\nu$ by means of first passage
calculation and replica trick and they found

\begin{equation}
\nu=\frac{1}{1+\beta}.\label{nu}
\end{equation}

\vskip 3 truecm
\begin{figure}
\begin{center}
\epsfxsize=0.6\hsize\epsfbox{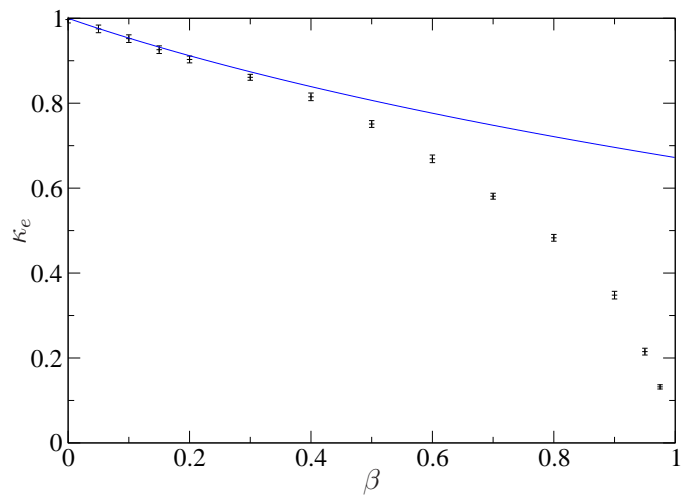}
\end{center}
\caption{Numerical value of $\kappa_e$ (plus) for
$\kappa_p=1$, compared with analytical results
(\ref{kubo}) (solid line).}
\label{kp1}
\end{figure}
We plot in figure \ref{nukp10} the value of the exponent $\nu$
fitted from simulations with (\ref{meandispa}) for different values
of $\kappa_p$. The results agree relatively well with equation (\ref{nu})
for temperatures below but close to the transition temperature.
However for very low temperature, the exponent is
significantly larger than that predicted by equation (\ref{nu}).
However we have verified that as the time of the simulation
is increased that the measured exponent appears to decrease (indeed this was the
case in \cite{tode} where analytical results and an effective trap model were available to
identify the correct exponent).

\vskip 3 truecm
\begin{figure}
\begin{center}
\epsfxsize=0.6\hsize\epsfbox{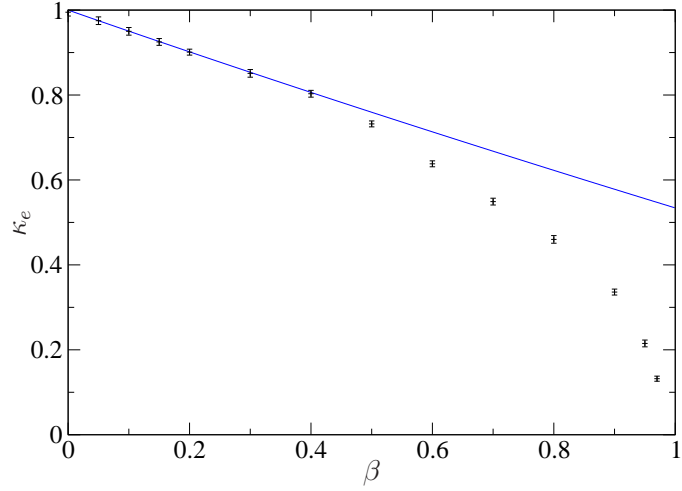}
\end{center}
\caption{Numerical value of $\kappa_e$ (plus) for
$\kappa_p=0.1$, compared with analytical results
(\ref{kubo}) (solid line).}
\label{kp0.1}
\end{figure}
\newpage
\eject
\begin{figure}
\begin{center}
\epsfxsize=0.5\hsize\epsfbox{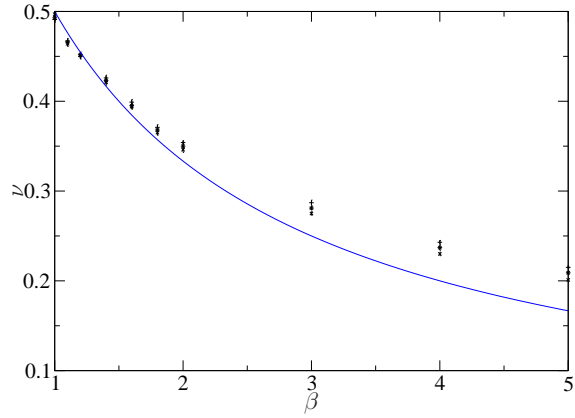}
\end{center}
\caption{Numerical value of $\nu$ (cross) for
$\kappa_p=10$, (stars) for $\kappa_p=1$  and
(plus) for $\kappa_p=0.1$ compared with analytical
results (\ref{nu}) (solid line) in the low
temperature phase.}
\label{nukp10}
\end{figure}

\section{Conclusions and discussion}

We have studied the dynamics of a dipole diffusing in a random
electrical field ${\bf E}$ derived from a quenched Gaussian
potential. In the adiabatic limit (where the dipoles adapt very
quickly to the local field), we showed that the coupled stochastic
equation can be reduced, up to the order $O(\kappa_x/\kappa_p)$, to
an effective Langevin equation for a particle diffusing in  an
effective potential with a spatially varying and anisotropic local
diffusivity $\kappa_{ij}$. In one dimension, we could compute
exactly the diffusion constant of this process and we found a
dynamical transition at finite temperature $\beta_c=1$ with a
crossover between a diffusive and a sub-diffusive regime. The
validity of the effective Langevin equation (\ref{eflang}) breaks
down for small $\beta$, but a high temperature treatment allowed us
to compute a Kubo formula of $\kappa_e$ for a given value of
$\kappa_p$. We  numerically checked our analytical predictions in
one dimension finding  good agreement and confirming the presence of
the transition. We also ran simulations far from  the adiabatic
regime and we found strong indications that the transition remains
at, or close to, the same $\beta_c$ as for the adiabatic limit.
Finally, we performed extensive numerical simulations in the low
temperature phase to compute the anomalous exponent $\nu$. We found
that it does not depend on the relative values of diffusion
constants of the particle's position and dipole moment and agrees
rather well with a calculation made for the adiabatic case
\cite{tode}.

In \cite{deto}, the authors showed that in higher dimensions, for
the adiabatic case case   $\kappa_p\to\infty$, that the same type of
dynamical transition is also present. It would be interesting to
investigate the dipole problem in higher dimensions to see if the
conclusions of our current study remain valid. It would also be
interesting to see if renormalization group type treatments or other
approximation schemes could be developed in order to obtain
analytical results in higher dimensions.

\appendix
\section{Derivation of the effective long-time Langevin equation for ${\bf x}$}

This appendix gives the detail of the calculation of the effective
long-time Langevin equation for ${\bf x}$. The first step is to formally
integrate the equation (\ref{leqp}) for the
variable ${\bf p}$ to obtain

\begin{equation}
p_i(t) = p_i(0) \exp(-{\beta\kappa_p\over \chi}t) + \int_0^t ds
\exp\left(-{\beta\kappa_p\over \chi}(t-s)\right) \kappa_p\beta E_i({\bf x}(s))
+ \sqrt{2\over \kappa_p}{\chi\over \beta} \zeta_i(t),\label{eqpi}
\end{equation}

where

\begin{equation}
\zeta_i(t) ={\kappa_p\beta\over \chi} \int_0^t ds
\exp\left(-{\beta\kappa_p\over \chi}(t-s)\right) \eta_{p_i}(s)
\label{eqcn}
\end{equation}

is a colored noise. At late times the first term of equation (\ref{eqpi})
is exponentially suppressed and can thus be dropped giving

\begin{equation}
p_i(t) = \int_0^t ds \exp\left(-{\beta\kappa_p\over \chi}(t-s)\right)
\kappa_p\beta E_i({\bf x}(s)) + \sqrt{2\over \kappa_p}{\chi\over \beta} \zeta_i(t).
\end{equation}

 Substituting this into equation (\ref{leqx}) yields

\begin{equation}
{dx_i\over dt} = \beta \kappa_x {\partial E_j \over \partial x_i}
\left[\int_0^t ds \exp\left(-{\beta\kappa_p\over \chi}(t-s)\right)
\kappa_p\beta E_j({\bf x}(s)) + \sqrt{2\over \kappa_p}{\chi\over \beta}
\zeta_j(t)\right]+\sqrt{2 \kappa_x}\eta_{x_i}.\label{eqfullx}
\end{equation}

To proceed further we will find an effective Fokker-Planck equation
corresponding to the stochastic equation (\ref{eqfullx}). We follow
the standard procedure of evaluating the change of an arbitrary function of $x(t)$:

\begin{eqnarray}
\langle {df({\bf x}(t))\over dt}\rangle &=& \int d{\bf x}
{\partial \rho({\bf x},t)\over \partial t} f({\bf x})\nonumber\\
&=&  \int d{\bf x} H \rho({\bf x},t) f({\bf x})  \nonumber \\
&=&  \int d{\bf x} \rho({\bf x},t) H^{\dagger}f({\bf x})
= \langle H^{\dagger}f({\bf x}(t)) \rangle \label{eqfp}
\end{eqnarray}

 where $\rho({\bf x},t)$ is the probability density for the process
${\bf x}$ at time $t$. The operator $H$ is the forward Fokker-Planck
operator and $H^{\dagger}$ is its adjoint commonly called the backward
Fokker-Planck operator.  We now write using the Stratonovich
prescription for white noise in stochastic calculus (where normal
differentiation applies) to give

\begin{eqnarray}
\langle {df({\bf x}(t))\over dt}\rangle  &=& \langle
{\partial\over \partial x_i} f({\bf x}(t))
{d{\bf x_i}(t)\over dt}\rangle \nonumber \\
&=& T_1 + T_2 + T_3.\label{opeall}
\end{eqnarray}

The first term is

\begin{equation}
T_1 = \sqrt{2\kappa_x} \langle \nabla_i f({\bf x}(t))
\eta_{x_i}\rangle,\label{ope1}
\end{equation}

which can be evaluated via Novikov's theorem as

\begin{equation}
T_1 = \sqrt{2\kappa_x} \langle {\partial^2 f\over
\partial x_i\partial x_j}\int_0^t ds\ {\delta x_j(t)\over \delta
\eta_{x_l}(s)}\langle \eta_{x_l}(s)\eta_{x_i}(t)\rangle\rangle
\label{ope1ev}
\end{equation}

which is true for any Gaussian noise. Now for a white noise
correlation, if we work with a symmetrised Dirac delta function
corresponding to the Stratonovich prescription, the above integral
picks up half the weight of the delta function. And we obtain

\begin{equation}
T_1 = \sqrt{{\kappa_x\over 2}} \langle
{\partial^2 f\over \partial x_i\partial x_j}
{\delta x_j(t)\over \delta \eta_{x_i}(s)}|_{s=t}\rangle.
\label{eqt11}
\end{equation}

The functional derivatives required in the above calculations can
be evaluated using a path integral formalism as in \cite{sasa}.
However they can also be obtained directly from the stochastic
equation as follows. Consider a general differential equation of the form

\begin{equation}
{dx_i\over dt} = W_i + B_{ij}\xi_j.\label{geneeq}
\end{equation}

In order to compute $\delta x_i(t)\over \delta\xi_j(s)$
one can define the process

\begin{equation}
{dx_i^{\epsilon}\over dt}=W_i + B_{ij}(\xi_j +\epsilon_j\delta(t-s)).
\label{eqfd1}
\end{equation}

Now from one of the standard definitions of the
functional integral we have

\begin{equation}
{\delta x_i(t)\over \delta \xi_j(s)} =
{\partial x_i^{\epsilon}(t)\over \partial \epsilon_j}|_{\epsilon=0}
.\label{def}
\end{equation}

Integration of equation (\ref{eqfd1}) in an infinitesimal
interval about $t=s$ yields

\begin{equation}
x^\epsilon(t^+) = x(t^-) + B_{ij} \epsilon_j,
\label{intxtp}
\end{equation}

as long as the $W_i$ do not depend on the derivative of $x$.
We see that in the case considered here the $W_i$ will be continuous
at $s=t$ and thus give no contribution to the functional derivative.
This thus yields

\begin{equation}
{\delta x_i(t)\over \delta \xi_j(s)} |_{s=t} = B_{ij}
.\label{Bij}
\end{equation}

Using this result we obtain

\begin{equation}
{\delta x_j(t)\over \delta \eta_{x_i}(s)}|_{s=t}
 = \sqrt{2\kappa_x}\delta_{ij},\label{resulBij}
\end{equation}

which upon substitution in equation (\ref{eqt11}) gives

\begin{equation}
T_1 = \kappa_x \langle {\partial^2\over \partial x_i \partial x_i}
 f({\bf x}_t)\rangle.\label{t1finl}
\end{equation}

This is the familiar Laplacian form arising for standard white noise.
The second term $T_2$ is given by

\begin{equation}
T_2 = \sqrt{2\kappa_x}\chi \sqrt{{\kappa_x\over \kappa_p}}\langle
{\partial E_j\over \partial x_i}{\partial f\over \partial x_i}\zeta_j(t)\rangle
.\label{ope2}
\end{equation}

Once again using Novikov's theorem we find

\begin{equation}
T_2 = \sqrt{2\kappa_x}\chi \sqrt{{\kappa_x\over \kappa_p}}\langle
{\partial\over \partial x_k}\left( {\partial E_j\over \partial x_i}
{\partial f\over \partial x_i}\right)\int_0^t ds{\delta x_k(t)
\over \delta \zeta_l(s)}\langle \zeta_j(t)\zeta_l(s)\rangle
.\label{ope2ev}
\end{equation}

The correlation function of the  Gaussian field $\zeta$ is easily
computed and is given for $t>s$ by,

\begin{equation}
\langle \zeta_i(t) \zeta_j(s)\rangle = \alpha^2 \int_0^s du
\ \exp\left(-\alpha (t+s-2u)\right),\label{corzeta}
\end{equation}

where $\alpha = \kappa_p\beta/\chi$. At large $t$ and $s$ this becomes

\begin{equation}
\langle \zeta_i(t) \zeta_j(s)\rangle = {\alpha \delta_{ij}\over 2}
\exp(-\alpha(t-s)).\label{corzetalargt}
\end{equation}

We now consider the action of this correlation function as an
operator on an arbitrary function $g(s)$ via the following operator
expansion

\begin{equation}
\int_0^t \ ds \exp\left(-\alpha s)\right) g(t-s) ={1\over \alpha}
\left[ g(t) -{1\over \alpha}{d\over dt}g(t) + O({1\over \alpha^2})
\right]\label{opeexp}
\end{equation}

and to leading order in $1/\alpha$ we thus obtain

\begin{equation}
T_2 =\sqrt{2\kappa_x}\chi\sqrt{{\kappa_x\over \kappa_p}}\langle
{\partial\over \partial x_k}
\left( {\partial E_j\over \partial x_i} {\partial f\over \partial x_i}\right)
{\delta x_k(t)  \over \delta \zeta_j(s)}|_{t=s} \rangle.\label{eqt21}
\end{equation}

Now repeating the argument leading to equation (\ref{resulBij})
we find that

\begin{equation}
{\delta x_k(t)  \over \delta \zeta_j(s)}|_{t=s} = \sqrt{2\over \kappa_p}
 \chi \kappa_x{\partial E_j\over\partial  x_k}\label{result2}
\end{equation}

and thus

\begin{equation}
T_2 =\kappa_x \chi^2\left({\kappa_x\over \kappa_p}\right)
\langle {\partial E_j\over \partial x_k} {\partial\over
\partial x_k} \left({\partial E_j\over \partial x_i}
{\partial f\over \partial x_i} \right)\rangle.\label{t2finl}
\end{equation}

One thus sees that the noise $\zeta$ is to leading order
white noise interpreted via the Stratonovich prescription.
The final term to evaluate is

\begin{equation}
T_3 = \beta^2 \kappa_x \kappa_p \langle {\partial f({\bf x}(t))\over
\partial x_i}{\partial E_j({\bf x}(t))\over \partial x_i}\int_0^t ds
\exp(-\alpha(t-s) )E_j({\bf x}(s))\rangle\label{ope3}
\end{equation}

and so one needs to compute terms of the form

\begin{equation}
I = \langle A({\bf x}(t))\int_0^t ds \exp(-\alpha(t-s) )
B({\bf x}(s))\rangle,\label{termI}
\end{equation}

bearing in mind that $\alpha$ is large. Now if
$\rho({\bf x},t)$ is the probability density for
the position ${\bf x}$ and $H$ the effective forward
Fokker Planck operator, we may write the above as

\begin{equation}
I = \int d{\bf x} B({\bf x}) \rho({\bf x}, s)
\exp(- (t-s)(\alpha-H^\dagger))A({\bf x}).\label{I2}
\end{equation}

The density $\rho$ obeys the Fokker Planck equation

\begin{equation}
{\partial \rho\over \partial t} = H\rho.\label{eqpev}
\end{equation}

The integration over $s$ can be evaluated by integrating
by parts, we have

\begin{eqnarray}
&&\int_0^t ds \ \rho({\bf x}, s) \exp(-(t-s)
(\alpha-H^\dagger))= \rho({\bf x},t)\left[ 1-
\exp(-t(\alpha-H^\dagger))\right](\alpha-H^\dagger)^{-1}\nonumber \\
&-&\int_0^\infty ds \ {\partial \rho({\bf x},s)\over
\partial s}\exp(-(t-s)(\alpha-H^\dagger))(\alpha-H^\dagger)^{-1}
.\label{intbypart}
\end{eqnarray}

Now because the eigenvalues of $H$ must be negative or
zero, we may neglect the second term in the square brackets
on the right hand side above. Another integration by parts yields

\begin{equation}
\int_0^t ds \ \rho({\bf x}, s) \exp(-(t-s)(\alpha-H^\dagger))
=\rho({\bf x},t) (\alpha-H^\dagger)^{-1}- \left
[H \rho({\bf x},t)\right](\alpha-H^\dagger)^{-2},\label{intbypart2}
\end{equation}

where we have used equation (\ref{eqpev}) in the second
term of the right hand side and the operator $H$ in this
term only acts on $\rho$ inside the square bracket. All
other operators act on the right and we can now expand
in powers of $1/\alpha$ to obtain

\begin{eqnarray}
\int_0^t ds \ B({\bf x}) \rho({\bf x}, s) \exp(-(t-s)
(\alpha-H^\dagger)) A({\bf x})=\nonumber \\
\rho({\bf x}, t)\left[ {A({\bf x}) B({\bf x}) \over \alpha}
 + B({\bf x}){H^\dagger\over \alpha^2} A({\bf x})-
{H^\dagger\over \alpha^2} A({\bf x})B({\bf x})\right]
.\label{expalpha}
\end{eqnarray}

Now putting all this together yields

\begin{eqnarray}
\langle H^\dagger f({\bf x})\rangle &=& \kappa_x \langle
{\partial^2\over \partial x_i \partial x_i} f({\bf x}_t)\rangle\nonumber\\
&+& \kappa_x \left({\kappa_x\over \kappa_p}\right)\chi^2\langle
{\partial E_j\over \partial x_k} {\partial\over \partial x_k}
\left({\partial E_j\over \partial x_i}{\partial f\over \partial x_i}
\right)\rangle\nonumber\\
&+& \kappa_x\beta\chi \langle {\partial f\over \partial x_i}
{\partial E_j\over \partial x_i}E_j +{\chi \over \kappa_p\beta}
E_j H^\dagger {\partial f\over \partial x_i} {\partial E_j\over
\partial x_i}-{\chi \over \kappa_p\beta} H^\dagger {\partial
f\over \partial x_i} {\partial E_j\over \partial x_i}E_j\rangle
.\label{alltogeth}
\end{eqnarray}

This can be written in the form

 \begin{eqnarray}
 && \langle H^\dagger f({\bf x})\rangle  = \langle H_0^\dagger
f({\bf x})\rangle \nonumber \\&+&
\chi^2 {\kappa_x\over \kappa_p}\left[ \langle \kappa_x{\partial
E_j\over \partial x_k} {\partial\over \partial x_k}
\left({\partial E_j\over \partial x_i}{\partial f\over \partial x_i}
\right) + E_j H^\dagger {\partial f\over \partial x_i}
{\partial E_j\over \partial x_i}
- H^\dagger {\partial f\over \partial x_i}
 {\partial E_j\over \partial x_i}E_j\rangle\right]
+O(({\kappa_x\over \kappa_p})^2),\nonumber \\
\label{eqhdagger1}
\end{eqnarray}

where

\begin{equation}
H_0^\dagger f= \kappa_x[ {\partial^2 f\over \partial x_i\partial x_i}
 + \beta\chi {\partial f\over \partial x_i}
{\partial E_j\over \partial x_i}E_j ].\label{hdagf}
\end{equation}

This is the only term remaining strictly in the limit
$\kappa_p\to \infty$ and corresponds exactly to a Langevin
particle in an effective potential $V$ given by equation
(\ref{adia}). Now keeping the
first term of  $O(\kappa_x/\kappa_p)$ we may write equation
(\ref{eqhdagger1}) as

\begin{eqnarray}
 && \langle H^\dagger f({\bf x})\rangle  = \langle
H_0^\dagger f({\bf x})\rangle \nonumber \\
&+& \chi^2 {\kappa_x\over \kappa_p}\left[ \langle
\kappa_x{\partial E_j\over \partial x_k} {\partial\over
\partial x_k} \left({\partial E_j\over \partial x_i}
{\partial f\over \partial x_i} \right) + E_j H_0^\dagger
 {\partial f\over \partial x_i} {\partial E_j\over \partial x_i}-
H_0^\dagger {\partial f\over \partial x_i} {\partial E_j\over
 \partial x_i}E_j\rangle\right] \nonumber\\
&=& \langle H_0^\dagger f({\bf x})\rangle -\kappa_x
{\kappa_x\chi^2 \over \kappa_p} \langle{\partial \over
 \partial x_k}\left( {\partial E_j\over \partial x_k}
{\partial E_j\over \partial x_i}{\partial f \over
\partial x_i}\right) + \beta\chi {\partial E_j\over \partial x_k}
{\partial E_j\over \partial x_i}{\partial f \over \partial x_i}
E_l {\partial E_l\over \partial x_k}\rangle.\label{Okxkp}
\end{eqnarray}

Finally the above may be written as

\begin{equation}
\langle H^\dagger f({\bf x})\rangle = \langle {\partial
\over \partial x_k}\left(\kappa_{ki}{\partial f \over
\partial x_i}\right) -\beta \kappa_{ki}{\partial V\over
\partial x_k}{\partial f\over \partial x_i}\rangle
,\label{hdagf2}
\end{equation}

with $\kappa_{ij}$ the spatially varying and anisotropic
diffusivity tensor given by (\ref{kij}) and $V$ the
potential defined by (\ref{eqV}).


\begin{thebibliography}{10}
\bibitem{comment}
We assume that the system is in equilibrium in a finite volume and
the subsequent late time diffusion constant is computed at times
such that the particle has diffused over a distance much larger than
the system's correlation length but much small than the system's
size.
\end{thebibliography}
\end{document}